\documentclass[useAMS,usenatbib,usegraphicx,newapa]{mn2e}

\title{Puoko-nui: A Flexible High-speed Photometric System}
\author[P.~Chote et al.]
       {P.~Chote,$^{1}$\thanks{Email: paul.chote@vuw.ac.nz}
        D.~J.~Sullivan,$^{1}$\thanks{Email: denis.sullivan@vuw.ac.nz},
		R.~Brown,$^{1}$
		\newauthor
		S.~T.~Harrold,$^{2}$
		D.~E.~Winget,$^{2}$
		D.~W.~Chandler$^{3}$\\
$^1$School of Chemical \& Physical Sciences, Victoria University of Wellington,
    P.O. Box 600, Wellington, New Zealand.\\
$^2$Department of Astronomy, University of Texas,
    Austin TX 78712, USA. \\
$^3$Meyer Observatory, Clifton TX 76634, USA.}

\begin{document}

\date{Accepted 2014 February 20}

\pagerange{\pageref{firstpage}--\pageref{lastpage}} \pubyear{2013}
\maketitle

\label{firstpage}

\begin{abstract}
We describe a portable CCD-based instrumentation system designed to efficiently
undertake high time precision fast photometry. The key components of the system
are (1)~an externally triggered commercial frame-transfer CCD, (2)~a custom
GPS-derived time source, and (3)~flexible software for both instrument control
and online analysis/display. Two working instruments that implement this design
are described. The New Zealand based instrument employs a Princeton Instruments
(PI) 1k$\times$1k CCD and has been used with the 1\,m telescope at Mt.\ John
University Observatory, while the other uses a newer 1k$\times$1k
electron-multiplying CCD supplied by PI and is based at the University of Texas
at Austin. We include some recent observations that illustrate the capabilities
of the instruments.
\end{abstract}


\section{Introduction}\label{section:introduction}

There are a number of interesting astronomical phenomena with timescales between
tens of milliseconds and tens of minutes. These include g-mode pulsations in
white dwarfs \citep[e.g.][]{winget08,fontaine08,althaus10} and p-mode
oscillations in the hot subdwarf B stars \citep[e.g.][]{fontaine06}. Both these
classes of objects exhibit photometric variability with periods
\mbox{$\sim10^{2}$ -- $10^{3}$\,s}.

Other phenomena are the eclipsing short period double-degenerate systems
\citep[e.g.][]{hermes12b}, \mbox{$\sim10^1 - 10^2$\,s}; planetary occultations,
\mbox{$\sim10^{-1}$\,s}; and optical pulsars \citep[e.g.][]{cocke69},
\mbox{$\sim10^{-2}$\,s}.

Our primary interests are with the pulsating white dwarfs, which are unstable to
g-mode pulsation in several instability strips that are a function of the
surface temperature and the dominant chemical atmospheric constituent. These
stars are also predicted to be unstable to p-mode pulsations with periods
$\sim 1$\,s, but any consequent luminosity variations have so far never
been observed \citep[e.g.][]{silvotti11, kilkenny14}.

An important aspect of high-speed photometry is accurate timing. The
pulsating white dwarfs, in particular, can have complicated pulsation
spectra with closely spaced frequencies that can require tens of hours of
continuous observation to properly resolve. This requires a clock
that remains sufficiently stable over an extended period during an acquisition
run, and also between successive days to allow multiple acquisition
runs to be combined for analysis. Furthermore, groups like the Whole
Earth Telescope \citep{nather90} coordinate observations from
different observatories around the globe in order to minimize data gaps,
thus extending the problem to multiple geographically distributed clocks.

This was a challenge in earlier decades, but has now been essentially solved by
sourcing time from the Global Positioning System (GPS) network. GPS disciplined
clocks which provide accurate time to $\mu$s or better are readily available
from several commercial manufacturers. The challenge in designing an instrument
is how best to integrate this time information into the overall system.
A common approach \citep[e.g,][]{dhillon07} is to operate the CCD asynchronously
and record the GPS-determined time at the start and/or end of an exposure.
An alternative approach is to use a GPS-disciplined clock to trigger CCD
readouts at known times. We use this approach in our Puoko-nui instruments.

The increasing availability of electron-multiply\-ing frame-transfer
CCDs makes it feasible to take photon-noise limited exposures at high
frame-rates. It is therefore important that the timing resolution is
high enough so as to not artificially limit the capabilities of the instrument.

\section{Instrument Overview}

The basic design for our system evolved from an earlier instrument, Argos
\citep{nather04}, which pioneered efficient CCD photometry of pulsating white
dwarfs. This development marked a significant departure from the multi-channel
photomultiplier-based instruments \citep[e.g.][]{kleinman96,sullivan00a}
that were previously used for these observations.

The key components of the Argos design philosophy were to use a commercial
frame-transfer CCD in combination with GPS-derived timing signals in order to
accomplish accurate time-series photometry as fast as 1\,Hz, and essentially
free from dead-time (readout) losses. An online display of the incoming light
curves to guide observers was also included. One time-series CCD photometer 
(Agile) that has successfully evolved from the Argos design has already been
described in the literature \citep{mukadam11}.

Argos had several limitations which made it unsuitable to directly copy when
creating Puoko-nui (which means ``big eye'' in NZ Maori). A major problem was
the concentration of all the required control functions along with the
acquisition and analysis software in one PC. Although this configuration had the
apparent advantage of simplicity, it created difficulties when changes were
required. In addition, largely as a result of the camera manufacturer providing
an interface via inadequately described binary code, the software required a
specific version of the Linux operating system kernel that did not operate well
with newer hardware.

In the Puoko-nui design we have separated the timing and control functions from
the acquisition and analysis operations by developing a microcontroller-based
unit that operates independently of the acquisition PC.
This unit uses GPS signals to maintain accurate time and output synchronised
CCD control signals; it communicates with the acquisition PC via USB. We have
also developed more flexible software packages for acquisition and analysis.
This increased flexibility made it relatively simple to extend support to a
newer and somewhat different CCD camera from the same manufacturer, which
required the acquisition and analysis software to run in a Microsoft Windows
environment. We will discuss both instrument flavours in this paper.

\begin{figure}
\centering
\includegraphics[width=\columnwidth]{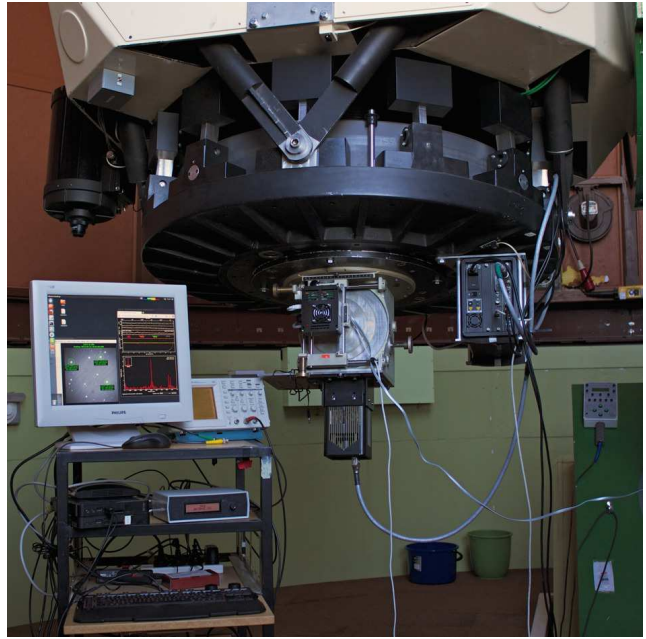}
\caption{The complete Puoko-nui South instrument mounted on the 1\,m~McLellan
telescope at Mt John University Observatory, Lake Tekapo, New Zealand.
\label{figure:instrumentpicture}}
\end{figure}

The original Puoko-nui (South) photometer, pictured in
Fig. \ref{figure:instrumentpicture}, is based in
NZ and has been operating in its current form since mid 2011; publications that
include observational data acquired with this instrument are listed in Section
\ref{section:results}. The second instrument (Puoko-nui North)
is based at the University of Texas at Austin and has been operational since
late 2012.

\begin{figure*}
\centering
\includegraphics{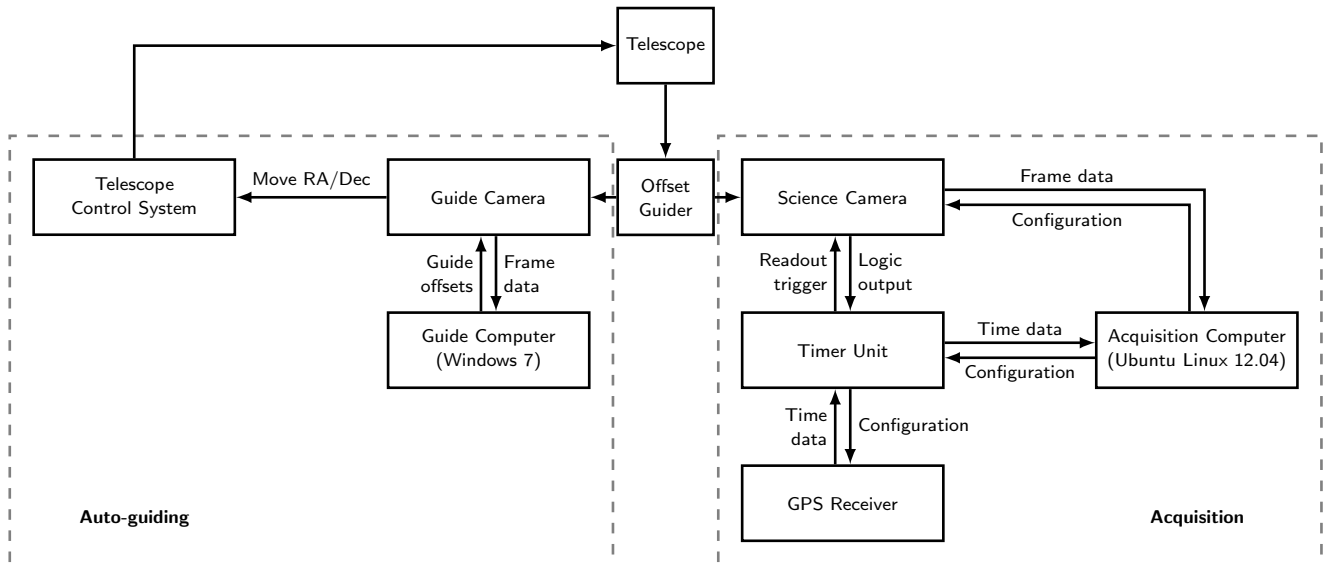}
\caption{A block diagram depicting the main components of Puoko-nui South.
Puoko-nui North mounts the science camera directly on the telescope, and
auto-guides using the science frames.\label{figure:instrumentblock}}
\end{figure*}

Fig.\ \ref{figure:instrumentblock} gives a schematic overview of the main
components of the Puoko-nui system, and they are explained in more detail in the
following sections.

\subsection{Science Cameras}\label{section:camera}

Both instruments use Princeton Instruments (PI) camera systems, containing E2V
back-illuminated $1024 \times 1024$\,px frame-transfer CCDs with
$13 \times 13\,\mu$m pixels. Puoko-nui South employs a PI MicroMax camera with a
USB data interface, while Puoko-nui North uses a PI ProEM camera with a gigabit
ethernet interface. The detailed specifications for the two cameras are shown in
Table~\ref{table:cameraspecs}.

The back-illumination of a CCD improves the quantum efficiency at shorter
wavelengths, which is particularly useful when observing the hot white dwarfs.
The use of a frame-transfer CCD further improves the efficiency by essentially
eliminating readout deadtimes.

The readout of a CCD (as compared with e.g.\ the CMOS sensor used more commonly
in consumer cameras) is a serial process, in which the charge from each pixel is
clocked across the sensor towards a single (or small number of) readout
registers. The readout process can take several seconds or more, during which a
standard CCD must be shuttered from incident light in order to prevent the
creation of additional extraneous charge in the pixels.

Frame-transfer CCDs contain twice the area of a regular sensor, with half of it
permanently masked as a readout storage region. The accumulated charge is
shifted to this region at the end of an exposure, allowing the time-consuming
final readout phase to occur in parallel with the next exposure.

This rapid shifting of charge between exposures removes the need for a
mechanical shutter, allowing the instrument to operate without moving parts.
When operated this way, there will be a small amount of image smearing caused by
photons impacting the CCD while the charge is being shifted into the storage
region, but given the very short transfer times this effect may only be
non-negligible at the shortest exposures. For the ProEM camera a full frame
transfer takes only $\sim1\,$ms, while it is $\sim 16\,$ms for the MicroMax
camera.

The ProEM camera also features an electron-multiplication (EM) readout port
\citep{mackay01}, which allows a pre-amplification of up to $1000\times$ to be
applied to the integrated image before it is measured. This can allow the
signal to overcome the readout noise (which is often the dominant noise source),
making it practical to acquire frames with a greater time resolution. The EM
readout mode also features faster readout rates, which correspondingly reduces
the minimum exposure length; 0.12\,s for the fastest full-frame readout at
10\,MHz, compared with 1.2\,s for the MicroMax at 1\,MHz. In both cases, the
readout time can be further reduced by reading out a smaller sub-area of the
CCD.

Both cameras are thermo-electrically cooled, with the MicroMax camera operating
at $-50^\circ$C, and the ProEM at $-55^\circ$C. The ProEM includes an additional
heat exchanger that can be connected to a liquid cooling system for increased
cooling efficiency.

The key feature that makes these cameras suitable for our purposes is that they
allow the frame transfer and subsequent readout to be triggered externally.
The internal oscillators used within the cameras were measured to be stable to
$\sim 1$ part in $10^5$, which is excellent for a single exposure, but this
accumulates to a total drift of as much as 50\,ms per hour.
The external trigger enables the exposure timing to be controlled by an external
clock, which can be disciplined against a reference time standard (e.g. GPS) to
provide consistent exposure triggers that remain aligned to the UTC-second
boundary over arbitrarily long run lengths.

\begin{table*}
\centering
\caption{A comparison of the science cameras in the two instruments.
The noise measurement for the ProEM's Electron Multiplication mode is given for
$1000 \times$ avalanche gain.\label{table:cameraspecs}}
\begin{tabular}{lccc}
\hline
\multicolumn{1}{l}{Parameter} & Puoko-nui South & \multicolumn{2}{c}{Puoko-nui North} \\
\hline
PI Model           & MicroMax             & \multicolumn{2}{c}{ProEM}\\
Connectivity       & USB 2.0              & \multicolumn{2}{c}{Gigabit Ethernet}\\
CCD Type           & E2V Back Illuminated & \multicolumn{2}{c}{E2V Back Illuminated EMCCD}\\
Active Area (px)   & $1024\times1024$     & \multicolumn{2}{c}{$1024\times1024$}\\
Mode               & Frame Transfer       & \multicolumn{2}{c}{Frame Transfer}\\
Pixel Size ($\mu$m)& 13                   & \multicolumn{2}{c}{13}\\
Vertical Clock Time ($\mu$s / row) & 15.2 & \multicolumn{2}{c}{1}\\
ADC Bit Depth (bits)            & 16      & \multicolumn{2}{c}{16}\\
Full Well Depth (ke$^-$)        & 110     & \multicolumn{2}{c}{120}\\
Dark Current (e$^-$/ px / s)   & 0.2      & \multicolumn{2}{c}{0.01}\\
Shutter            & None                 & \multicolumn{2}{c}{Electro-mechanical}\\
Software API       & PVCAM (32-bit Linux) & \multicolumn{2}{c}{PICAM (64-bit Windows)}\\
\cline{3-4}
Readout Type       & Low Noise            & Low Noise & EM\\
\cline{3-4}
Readout Rates (MHz)& 0.1, 1               & 0.1, 1, 5 & 5, 10 \\
Noise (e$^-$ rms)  &5.0 -- 13    & 3.2 -- 13  & 0.03 -- 0.04 \\
Gain (e$^-$/ ADU) & 1.0 -- 4.5  & 0.7 -- 3.4 & 2.1 -- 11.4\\
EM Gain            & --          & --         & 1 -- 1000\\
\hline
\end{tabular}
\end{table*}

\subsection{Timing hardware}
\begin{figure}
\centering
\includegraphics[width=\columnwidth]{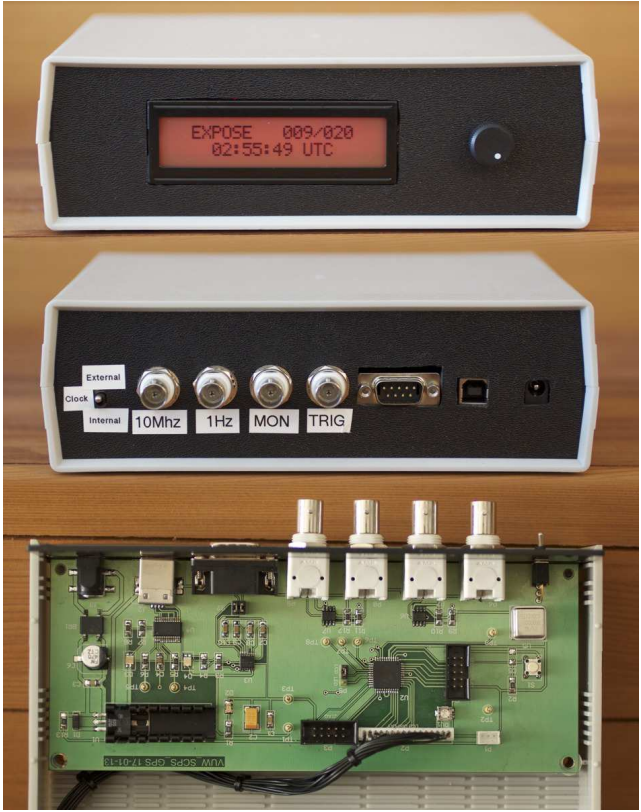}
\caption{The Puoko-nui timer unit is built around a stand-alone microcontroller,
allowing it to operate independently during an acquisition sequence. Time and
exposure information is sent to the Acquisition PC via a standard USB 2.0
connection.\label{figure:timerpicture}}
\end{figure}

Precise exposure timing is provided by a custom timer/counter unit, which is
pictured in Fig. \ref{figure:timerpicture}. The unit is built around an Atmel
AVR microcontroller, and communicates with the acquisition PC via USB. The USB
connectivity, paired with the USB/Ethernet cameras allows the instrument to be
controlled using a standard PC laptop or small-form-factor `netbook' computer
without requiring any additional hardware.

The primary function of the timer unit is to process time signals from an
external GPS disciplined clock (currently supporting the Trimble Thunderbolt and
a legacy Magellan OEM receiver) to manage the exposure timing. The GPS time
signals are provided in the form of 1\,Hz and 10\,MHz pulse trains via coaxial
cables, and a RS232 serial stream of ASCII-coded absolute time information.

The unit features two timing modes, which operate by counting pulses on either
the 10\,MHz pulse input (high-resolution mode) or the 1\,Hz pulse input (low
resolution mode). In both modes the unit counts pulses up to the desired
exposure length, and then outputs a frame-transfer trigger to the camera. The
GPS time associated with each trigger is sent to the acquisition PC to be
written into the frame header data.

Externally triggering the individual frame transfers simplifies the process of
integrating precise timing with a commercial camera system. This separation of
timing and acquisition was an important factor in the rapid development of
Puoko-nui North, and allows for additional camera systems to be supported in the
future with only minimal changes.

The two timing modes are compared in Table \ref{table:timermodes}.
The high-resolution mode is preferred, as it contains additional checks for
potential timing discrepancies. The low-resolution mode remains available as a
fallback for acquiring exposures longer than 65 seconds, or when operating using
our Magellan GPS receiver, which does not conveniently provide a 10\,MHz signal
in its current form.

\begin{table}
\centering
\caption{The timer unit features two operating modes. The high-resolution mode
supports trigger periods as short as $1\,$ms, but in practice the camera readout
speed limits exposures to $>5\,$ms and telescope-size currently limits useful
white-dwarf observations to $>1\,$s for Puoko-nui South.\label{table:timermodes}}
\begin{tabular}{lcc}
\hline
\multicolumn{1}{c}{Mode} &   High-res.      &  Low-res. \\
\hline
Exp. range               & 1\,ms -- 65.5\,s & 1\,s -- 18.2\,h \\
Exp. resolution          & 1\,ms            & 1\,s \\
Trigger delay            & $<1\,\mu$s       & 18\,ms\\
Trigger instability      & $<10\,\mu$s      & $<10\,\mu$s\\
Required inputs & 1\,Hz, 10\,MHz, RS232     & 1\,Hz, RS232\\
\hline
\end{tabular}
\end{table}

The trigger delay figure in Table \ref{table:timermodes} was
obtained by comparing the 1\,Hz GPS pulse train against the generated trigger
pulses using an oscilloscope. The camera response time between the trigger and
frame transfer beginning was similarly measured at $ < 1\,\mu$s. In the case of
the high-resolution mode, the phase of the trigger pulse can be advanced or
delayed in order to minimize the difference against the UTC-second aligned 1\,Hz
input. We have explicitly verified the absolute timing accuracy of Puoko-nui
South to better than 20\,ms by imaging the time display of the older
three-channel photometer master clock, which was set using an independent GPS
receiver. Observations of the Crab Pular using Puoko-nui North
(see Section \ref{section:results}) further verifies the timing accuracy to
within a few ms, limited by the precision of our software reduction pipeline.
The trigger instability figure is calculated based on the possibility of a
hardware interrupt being delayed while another is serviced, which may introduce
a delay of up to 100 clock cycles (10\,$\mu$s). This situation is rare, and the
measured jitter usually remains below 1\,$\mu$s.

The shortest possible exposure time is limited by the camera readout
rate (which in itself depends on factors such as the readout window parameters),
but the exposure time will be limited in most situations by the accumulation of
sufficient photons from the faint targets. For maximum flexibility, we wanted
to ensure that the timing system did not impose any additional constraints,
particularly for a portable instrument that may have the opportunity to be used
on telescopes larger than the original design specified.

A secondary function of the timer is to monitor the camera readout status via a
logic output. This is necessary to work around a bug in the closed source
camera firmware and \textsc{pvcam} library for the MicroMax camera: terminating
an exposure sequence mid-readout can corrupt the internal state of the
`black-box' software and cause subsequent exposure sequences to be corrupted.
The software API does not allow the camera status to be queried, and so a
hardware work-around was implemented using the timer to monitor the programmable
logic output on the camera to determine its current state.

The USB connection between the timer and acquisition PC is configured to provide
a 9600\,baud serial communication channel. We have chosen this limited data
rate (USB allows speeds up to 3\,Mbaud) in part to match the GPS receivers.
This allows the timer (when operating in its `relay' mode) to transparently
forward data between the GPS receiver and acquisition PC so that other software
on the PC can interact with it. The limited data rate is sufficient for normal
operation with exposure lengths greater than 0.1\,s. Shorter exposures are
supported by reducing the number of trigger timestamps sent to the acquisition
PC (sending only 1 in N exposures, where N is chosen to keep the data transfer
similar to 0.5\,s exposures). The intermediate timestamp values are synthesised
by the acquisition PC.

\subsection{Acquisition software}\label{section:software}

The acquisition and control software is written primarily in the C programming
language and is portable across 32 and 64 bit versions of Linux, Windows and
Mac~OS~X. In practice, the availability of (proprietary) camera drivers
currently restricts operation to 32 bit Linux (using Ubuntu 12.04) for
Puoko-nui South, and 64-bit Windows~7 for Puoko-nui North.

\begin{figure}
\centering
\includegraphics{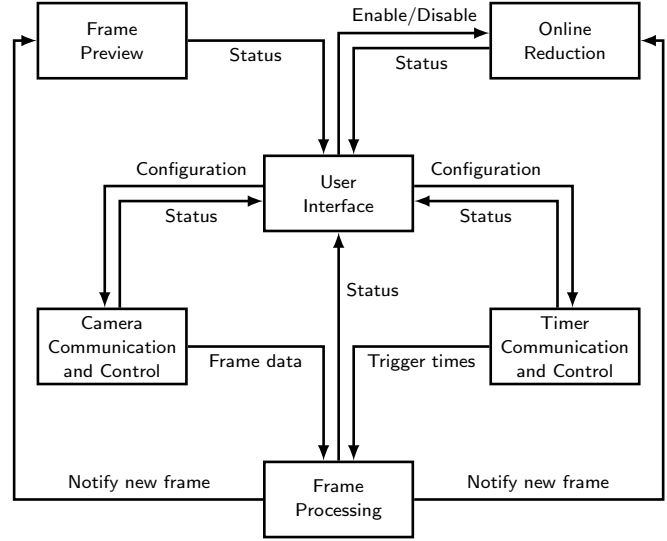}
\caption{A block diagram depicting the key software threads in the
Puoko-nui acquisition software.\label{figure:acquisitionblock}}
\end{figure}

The acquisition software performs three main tasks: configuring the acquisition
parameters, matching frame data with timestamps during acquisition, and
notifying the user of the current hardware status. The acquisition software
is divided into a number of concurrent threads that communicate asynchronously. 
The components and main information flow are shown in
Fig.\ \ref{figure:acquisitionblock}.

\begin{figure}
\centering
\includegraphics[width=\columnwidth]{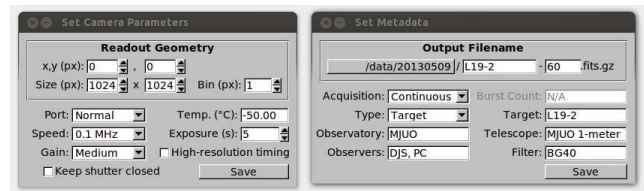}
\caption{A screen capture showing the acquisition configuration and frame
metadata available in the Puoko-nui acquisition software.\label{figure:instrumentconfig}}
\end{figure}

Fig.\ \ref{figure:instrumentconfig} shows the run configuration parameters and
the customisable information fields that are saved into the frame headers.
The camera and timer are configured with these parameters when the user starts
an acquisition run, and the software then becomes responsive to incoming frame
data and trigger timestamps.

Frame data from the camera and trigger timestamps from the timer are received
from their respective hardware interfaces, and passed to a processing thread.
The processing thread performs a basic validation check to ensure that the
frame being associated with each timestamp is correct (the time that the frame
was received minus the readout time should match the trigger timestamp to within
a small margin of error), and then saves the frame and associated metadata to
disk as a compressed FITS image.

The Puoko-nui South instrument can acquire arbitrarily long acquisition
sequences, limited only by disk space. The faster readout rates supported by
Puoko-nui North can result in the situation where the data rate from the camera
exceeds the disk write rate; frames will be buffered in memory until the backlog
of frames have been purged to disk. This could be improved by installing a
faster solid-state hard disk, however this is unnecessary for the $\sim$ second
exposures used for our primary white dwarf targets.

The most recently acquired frame is displayed using the SAOImage DS9 software,
and can be overlaid with information on one or more selected comparison stars.
This information is calculated on a per-frame basis independently of the online
reduction, and can be used for auto-guiding the telescope
(see Section \ref{section:autoguiding}).

\begin{figure}
\centering
\includegraphics[width=\columnwidth]{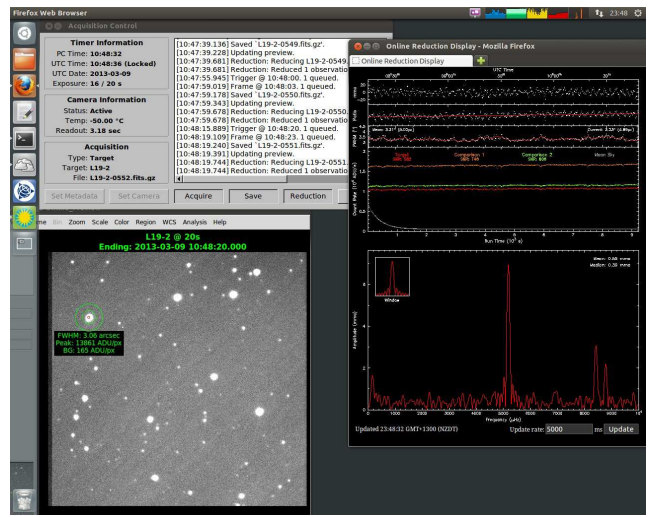}
\caption{A screen capture showing the online display produced by the acquisition
and analysis software during an observation run. The target for these
observations is the hydrogen atmosphere pulsating white dwarf L19-2. The star
selected in the frame display is a (non-variable) comparison star used to
monitor observing conditions.\label{figure:softwarepicture}}
\end{figure}

\subsection{Online Reduction}
We have developed reduction software, called \textsc{tsreduce}, that performs
real-time analysis on the acquired CCD frames, and features offline analysis for
generating final data reductions. Synthetic aperture photometry techniques are
used to calculate light curves of the target and selected comparison stars.
The effects of cloud and atmospheric extinction changes are compensated for by
taking the ratio of target and comparison lightcurves, and then any remaining
long period trends in the relative lightcurve (resulting from differential
colour extinction effects for example) can be minimized by fitting a low order
polynomial. This latter procedure is appropriate for studies of the pulsating
white dwarf stars, and can be disabled.

The resulting data corresponds to fractional intensity changes in the target
star, and are displayed as a plot of milli-modulation intensity
(10\,mmi = 1\% change) versus time. The discrete Fourier transform (DFT) of the
data is calculated in amplitude units (which has a more direct physical
connection with the data than the more traditional power units) and displayed
versus frequency along with the DFT window function.

The reduction display also includes the raw lightcurve plots and an estimate of
the mean full-width at half-maximum of the point spread functions of the
selected stars. This information enables the observer to monitor atmospheric
conditions and telescope focus.

The online reduction has a simple configuration procedure where the user selects
the target and comparison stars and surrounding annuli for measuring the
background intensity. The aperture size is estimated using a curve of growth
technique on the first frame in the exposure sequence. The aperture size is not
varied between frames, and so this single-frame estimate is rarely optimal, but
is sufficient for the purposes of an online display.

The offline reduction routines allow an optimized final reduction to be
calculated by determining the optimum aperture size for each data set, removing
the effect of the Earth's orbital motion from acquisition timestamps (using the
excellent \textsc{sofa} library \citep{sofa10}), and combining multiple data
sets into a standard output format for analysis. A set of offline analysis
routines simplify the identification of frequency modes and estimation of noise
limits. The reductions generated by \textsc{tsreduce} compare favorably with
other reduction codes such as \textsc{maestro} \citep{dalessio10},
\textsc{ccd\_hsp} \citep{kanaan02}, and \textsc{wqed} \citep{thompson09}.

Fig.\ \ref{figure:softwarepicture} shows a screen capture of the acquisition PC
mid-way through a typical observing run. The display features three windows
showing the online reduction plots, the latest acquired frame, and the
instrument control/status window.

\subsection{Auto-guiding}\label{section:autoguiding}

The frame preview display provides star positions which can be read by an
external program to enable auto-guiding via the science frames. This
functionality has been used when operating Puoko-nui North on the 2.1\,m Otto
Struve Telescope at McDonald Observatory in West Texas.

Puoko-nui South uses a separate dedicated auto-guiding system. The PI camera is
mounted on the McLellan 1\,m telescope at Mt John University Observatory (MJUO)
using the offset guiding mount adapted from the earlier VUW 3-channel photometer
\citep[see][]{sullivan00a}. This mount contains a 45$^\circ$ mirror with a hole
that allows the central field of view to enter the primary camera. The annulus
outside the primary field of view is redirected to a secondary camera (an SBIG
ST402ME) mounted on a two-dimensional slide mechanism, which allows a nearby
star to be independently monitored for auto-guiding purposes. The SBIG camera
features a hardware interface that is connected to the telescope control system,
and the packaged software (which runs on a separate PC) features a dedicated
auto-guiding mode.

The main advantage of this arrangement is that the offset guider configuration
has access to a much larger field of view than the primary camera, thus ensuring
access to an adequately bright guide star in the often sparse WD fields.

\section{Some Results}\label{section:results}

Observational data obtained using Puoko-nui South that have already appeared in
print include observations of the dwarf nova GW\,Librae \citep{szkody12,chote13a},
the helium atmosphere pulsating white dwarf EC\,04207$-$4748 \citep{chote13b},
and the eclipsing double-WD system J0751 \citep{kilic13}.

We have undertaken a number of observations using both instruments primarily to
ascertain and verify their inherent timing capabilities.

Puoko-nui North in combination with the 2.1\,m telescope at McDonald Observatory
was used to observe the extremely stable hydrogen atmosphere pulsating white
dwarf G117$-$B15A. The impact of evolutionary cooling on this star results in a
gradual increase in the 215\,s mode period (d$P$/d$t$ $\sim 10^{-15}\,$ s\,s$^{-1}$),
which has been detected as a parabolic trend in the observed minus calculated
(O$-$C) phase plot over a long observation baseline \citep[see][]{kepler05a}.

\begin{figure}
\centering
\includegraphics{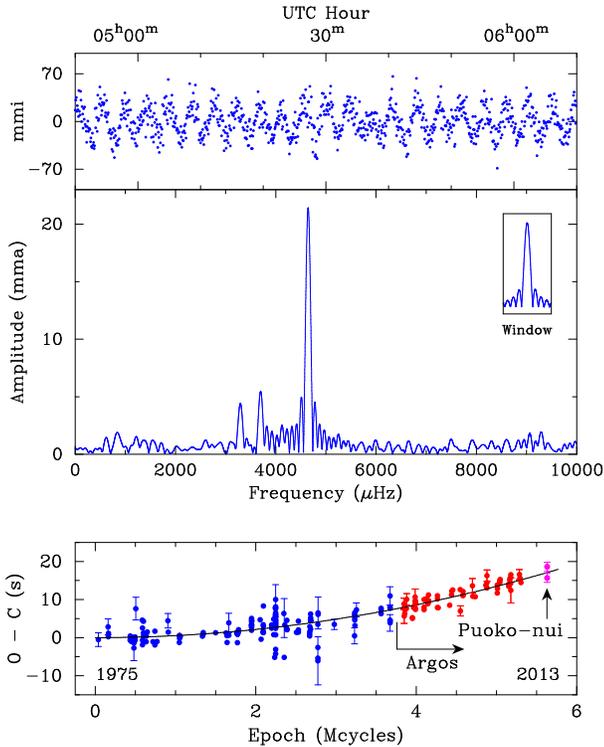}
\caption{The top two panels depict a short lightcurve segment (in
milli-modulation intensity units) and a discrete Fourier transform (in
milli-modulation amplitude units) from an observation run on the extremely
stable DAV white dwarf pulsator G\,117$-$B15A. Data from two runs were acquired
in April 2013 using the Puoko-nui North instrument attached to the 2.1\,m
telescope at McDonald Observatory. The bottom panel compares the measured phases
of the dominant 215\,s pulsation with the nearly four decades of archival data.
Only a representative subset of the measurement uncertainties have been included
in the plot to minimise clutter. These new phase measurements are consistent
with the theoretical trend (solid line) to within the measurement uncertainties.
As well as extending the O$-$C phase data for this pulsator, the measurements
also provide a check on the photometer's absolute timing accuracy (see text and
\citet{kepler05a} for more details).\label{figure:g117}}
\end{figure}

Fig.\ \ref{figure:g117} depicts an excerpt of these observations and, in
particular, the bottom panel compares our April 2013 O$-$C phase measurements
with the nearly four decades of archival data on this star. This plot provides a
simple check of the absolute timing accuracy of the instrument: any systematic
time offsets would appear as a vertical offset in the plot. This technique has
been used in the past to identify timing issues with the Argos photometer.
Our measurements agree with the calculated phase with an uncertainty of
$\pm 1.5$\,s [Kepler, private communication]. Alternatively, the new
observations can be viewed as extending the O$-$C measurements that detect the
impact of evolutionary cooling on this star. Referring to
Fig.\ \ref{figure:g117}, it is interesting to note the general improvement in
measurement precision that was obtained by moving from photomultiplier
instruments (blue points) to frame transfer CCD measurements (Argos, red points)
in 2002. Puoko-nui measurements continue this.

\begin{figure}
\centering
\includegraphics{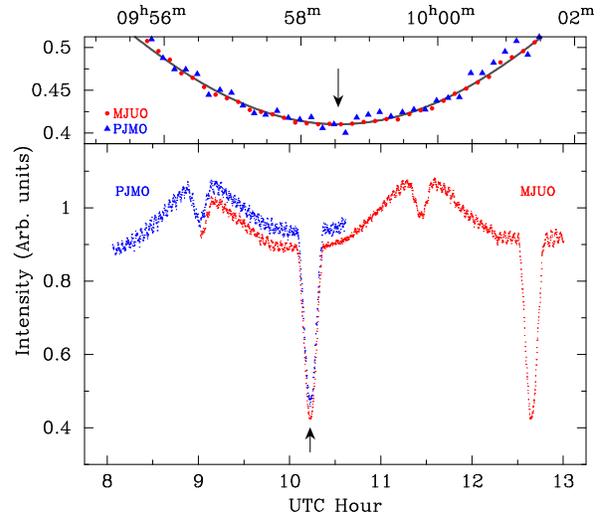}
\caption{Coordinated and concurrent observations of the near-equatorial short
period eclipsing binary PG\,1336$-$018 \citep{kilkenny03}, obtained in April
2013 using the Puoko-nui North photometer attached to the 0.6\,m robotic
telescope at Meyer observatory (PJMO), near Clifton, Texas and the Puoko-nui
south system with the 1\,m telescope at MJUO in NZ. The data in the bottom panel
have been offset vertically for clarity. The top panel shows a quartic fit to
the eclipse, and demonstrates that the data are consistent to within the
photometric uncertainties (given by the point size for the MJUO data, and a
factor of two larger for PJMO).\label{figure:pg1336}}
\end{figure}

Concurrent observations of the short period eclipsing binary PG\,1336$-$018
\citep{kilkenny03} using the two instruments provided a further test of the
relative timing integrity of both systems, and are shown in
Fig.\ \ref{figure:pg1336}. The individual timings of the primary eclipses are
consistent within the photometric uncertainties, which translates to a few
seconds.

\begin{figure}
\centering
\includegraphics{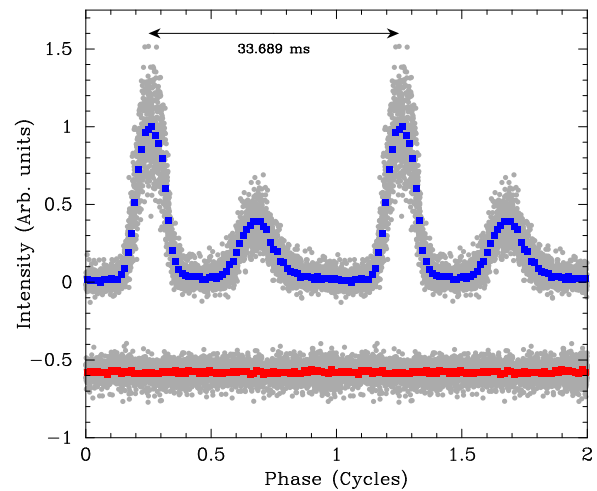}
\caption{Observations of the Crab pulsar demonstrate the high-speed capabilities
of the Puoko-nui North photometer. These data, consisting of a
sequence of 5000 5\,ms exposures, were obtained using the 2.1\,m telescope at
McDonald Observatory in November 2013. The plot shows the intensity of the
pulsar and a nearby comparison star (grey dots), folded at the primary period of
33.689\,ms. This is overlaid with the binned intensity for the pulsar (top,
blue) and comparison (bottom, red). The data have been repeated for a second
cycle, and the comparison star intensity offset vertically by -0.75 units to
improve clarity. \label{figure:crabpulsar}}
\end{figure}

The high-speed capability of Puoko-nui North was demonstrated by observing the
33\,ms optical pulsation of the Crab pulsar in November 2013. A sequence of 5000
5\,ms exposures were acquired using the 2.1\,m McDonald telescope; these short
exposures were achieved using the 10\,MHz electron-multiplying readout mode to
digitize a $336 \times 128$\,px region of the CCD with $8 \times 8$ binning. The
resulting $42 \times 16$\,px image covered a $30^{\prime\prime} \times
12^{\prime\prime}$ region surrounding the pulsar.
In Fig. \ref{figure:crabpulsar} we present lightcurves of the pulsar and a
nearby comparison star folded on a period of 33.689\,ms, which was determined
from a DFT of the pulsar data.  The phase of the main peak was compared with the
Jodrell Bank radio ephemeris \citep{lyne93}, and found to agree to within
2.5 $\pm$ 5\,ms.  This relatively large uncertainty comes from numerical
precision limitations in our data reduction pipeline, which is not currently
optimized for sub-ms accuracy BJD time conversions.

\section{Discussion}

The two CCD photometers that we describe here have a common GPS timing and
instrument control functionality, and they offer a number of distinct advantages
for undertaking the demanding task of obtaining precision high-speed time-series
photometry.

Both instruments are sufficiently portable that they can be readily transported
using airline checked luggage facilities (allowing for the associated weight
limitations). The NZ system is regularly transported this way between its
Wellington home base and Mt John University Observatory in the South Island.
However, transporting the mechanical offset guider box that is an `optional
extra' part of Puoko-nui South adds a complicating layer. The Puoko-nui North
system is even more portable as the newer PI ProEM camera has all the
electronics housed in one unit.

The flexibility of our photometer design makes it relatively simple to integrate
with other externally triggered CCD cameras or GPS receivers. In addition,
because most of the instrument behavior is defined in software, it would be
straightforward to add additional capabilities to e.g. acquire synchronous
photometry, read multiple `windows' within the CCD frame, or
support a filter wheel in order to carry out automated multi-colour photometry
(we have no immediate plans to do this however).
Similarly, the online reduction display can be easily modified to show the
information most relevant to the target: e.g. the DFT is not useful for many
types of time-series photometry, and can be disabled.

Except for the proprietary PI camera interface code, all other software is open
source. This includes the C language coding for both the timer firmware and the
acquisition and control PC. This software and the hardware schematics for the
timer are available on request.

\section*{Acknowledgements}
We would like to thank S.O. Kepler for providing the G\,117$-$B15A archival
phase and fit data.

PC and DJS thank the NZ Marsden Fund for financial support and the University of
Canterbury for the allocation of telescope time.

STH, DEW, and DWC gratefully acknowledge the Central Texas Astronomical Society
for the remote use of Meyer Observatory, McDonald Observatory for the use of the
2.1\,m and 0.9\,m telescopes, and funding support from McDonald Observatory and
the Longhorn Innovation Fund for Technology.

We also thank the anonymous referee whose thoughtful comments led to an
improvement in the quality of the paper.

\bibliographystyle{mn2e}
\bibliography{chote13}

\label{lastpage}
\end{document}